\documentclass{emulateapj}
\usepackage[latin1]{inputenc}
\usepackage{amsmath}
\usepackage{amsfonts}
\usepackage{amssymb}
\usepackage{graphicx}
\usepackage{gensymb}
\begin{document}

\title{Spatially resolved eastward winds and rotation of HD\,189733\MakeLowercase{b}}
\author{Tom Louden\altaffilmark{1} and Peter J.\ Wheatley\altaffilmark{1}}
\altaffiltext{1}{Department of Physics,
University of Warwick,
Coventry, CV4 7AL, UK}

\email{t.m.louden@warwick.ac.uk}

\begin{abstract}
We measure wind velocities on opposite sides of the hot Jupiter HD\,189733b by modeling sodium absorption in high-resolution HARPS transmission spectra. Our model implicitly accounts for the Rossiter-McLaughlin effect, which we show can explain the high wind velocities suggested by previous studies. Our results reveal a strong eastward motion of the atmosphere of HD\,189733b, with a redshift of $2.3^{+1.3}_{-1.5}$\,km\,s$^{-1}$ on the leading limb of the planet and a blueshift of $5.3^{+1.0}_{-1.4}$\,km\,s$^{-1}$ on the trailing limb. These velocities can be understood as a combination of tidally locked planetary rotation and an eastward equatorial jet; closely matching the predictions of atmospheric circulation models. Our results show that the sodium absorption of HD\,189733b is intrinsically velocity broadened and so previous studies of the average transmission spectrum are likely to have overestimated the role of pressure and thermal broadening.
\end{abstract}


\section{INTRODUCTION}\label{sec-intro}

Detections of exoplanetary winds to date have used high-resolution transmission spectra constructed from entire planetary transits to measure the average Doppler shift of atmospheric absorption around the planetary limb. In the hot Jupiter HD\,209458b a study with the CRyogenic high-resolution InfraRed Echelle Spectrograph (CRIRES) revealed a blue-shift of 2 $\pm$ 1\,km\,s$^{-1}$ in the average line profile of carbon monoxide absorption \citep{Snellen2010}. While in HD\,189733b a high-resolution optical study with the High Accuracy Radial Velocity Planet Searcher (HARPS) found a blue-shift of 8 $\pm$ 2\,km\,s$^{-1}$ in the atomic sodium doublet \citep{Wyttenbach2015} (although in this paper we show this value is too high). Blue-shifted absorption in the average transmission spectra indicates net wind flows from the dayside to the night side of the planets. 

Measurements of the average Doppler shift, however, cannot test atmospheric circulation models of hot Jupiters, which predict very different velocities on either side of the planet. A combination of tidally locked planetary rotation and an eastward equatorial jet \citep{Showman2002, Rauscher2010, Perna2010} are expected to result in red-shifted absorption on the leading limb of the planet and blue-shifted absorption on the trailing limb \citep{Showman2013, Miller2012}. The equatorial jet is believed to arise from planetary scale Rossby waves interacting with the mean atmospheric flow \citep{Showman2011}, and it is likely to be a common feature of heavily irradiated close-in planets \citep{Showman2015}, controlling the redistribution of heat from the permanent dayside to the night side of the planet. While not detected directly, the existence of an eastward jet is supported by the observation that the hottest point of the atmosphere of HD\,189733b is offset east of the sub-stellar point by 30\degree  \citep{Knutson2007, Knutson2008}.

In this paper we spatially resolve the rotation of the planetary atmosphere using a time resolved model of sodium absorption during the planet transit.

\section{Observations}\label{sec-intro}

We analysed data from a transit observation of HD\,189733b made with the HARPS instrument \citep{Mayor2003} mounted on the European Southern Observatory's 3.6m telescope at La Silla, Chile. We used data from ESO programme 079.C-0127(A) taken on the night of 28 August 2007, for which the planetary sodium lines have been shown to be free of systematic contamination by \citet{Wyttenbach2015} (false-alarm probability of $<0.01\%$). HD\,189733 was observed for a period of 4 hours, with 40 spectra taken in total, covering the two-hour transit with an additional one hour on either side for baseline comparison. We accessed the reduced spectra and data products through the ESO Archive and extracted the order from the echelle spectrum that contains the sodium feature (585.0 - 591.6 nm). The set-up of the HARPS spectrograph yields a resolution of 115,000 \citep{Mayor2003}. As well as the sodium absorption \citep{Wyttenbach2015}, these data have been used previously to measure the stellar rotation \citep{Cameron2010}, the Rossiter-McLaughlin effect \citep{Triaud2009} and to detect Rayleigh scattering in the planetary atmosphere \citep{Gloria2015}.

The spectra were aligned in a common velocity reference frame by removing both the barycentric velocity at the time of each spectrum as recorded by the HARPS pipeline, and the line-of-sight velocity of the star using the system orbital parameters \citep{Triaud2009}. This ensured that misalignment of the stellar line profiles did not introduce spurious features to the transmission spectra. 

Interstellar sodium absorption is tends to be negligible for stars within 50 pc \citep{Welsh1994} and we did not detect an interstellar component for HD 189733b, which is a relatively nearby star at only 19.5 pc. However, weak absorption features in the Earth's atmosphere were apparent in the raw spectra, and we took steps to remove the time varying component of the telluric absorption. All the science frames were combined into a single very high signal to noise spectrum, and this master spectrum was then subtracted from each individual frame, exposing the sources of variability. We created a model telluric spectrum from a detailed line list in the region surrounding the sodium doublet, which includes the relative strengths of all lines \citep{Lundstrom1991}. The telluric template spectrum was then fitted to each difference spectrum, with the FWHM, velocity offset and absolute strength of the spectrum as the free parameters. It was found that the quality of the fit could be improved by allowing atmospheric water lines to have a separate scaling factor. This model was then subtracted from the science frames, removing the variability in the telluric lines. 

The sodium absorption features in the Earth's atmosphere are weak in comparison to water features, and are offset from the stellar spectrum by the barycentric velocity of 9\,km\,s$^{-1}$ during the observation window. They are therefore unable to mimic a planetary transmission signal in our data.

Finally, the individual spectra were normalised in order to remove the absolute depth of the planetary transit, leaving only the relative transmission as a function of wavelength.

\section{Transit Model}\label{sec-method}

\begin{figure}
\begin{center}
\includegraphics[width=1.0\linewidth]{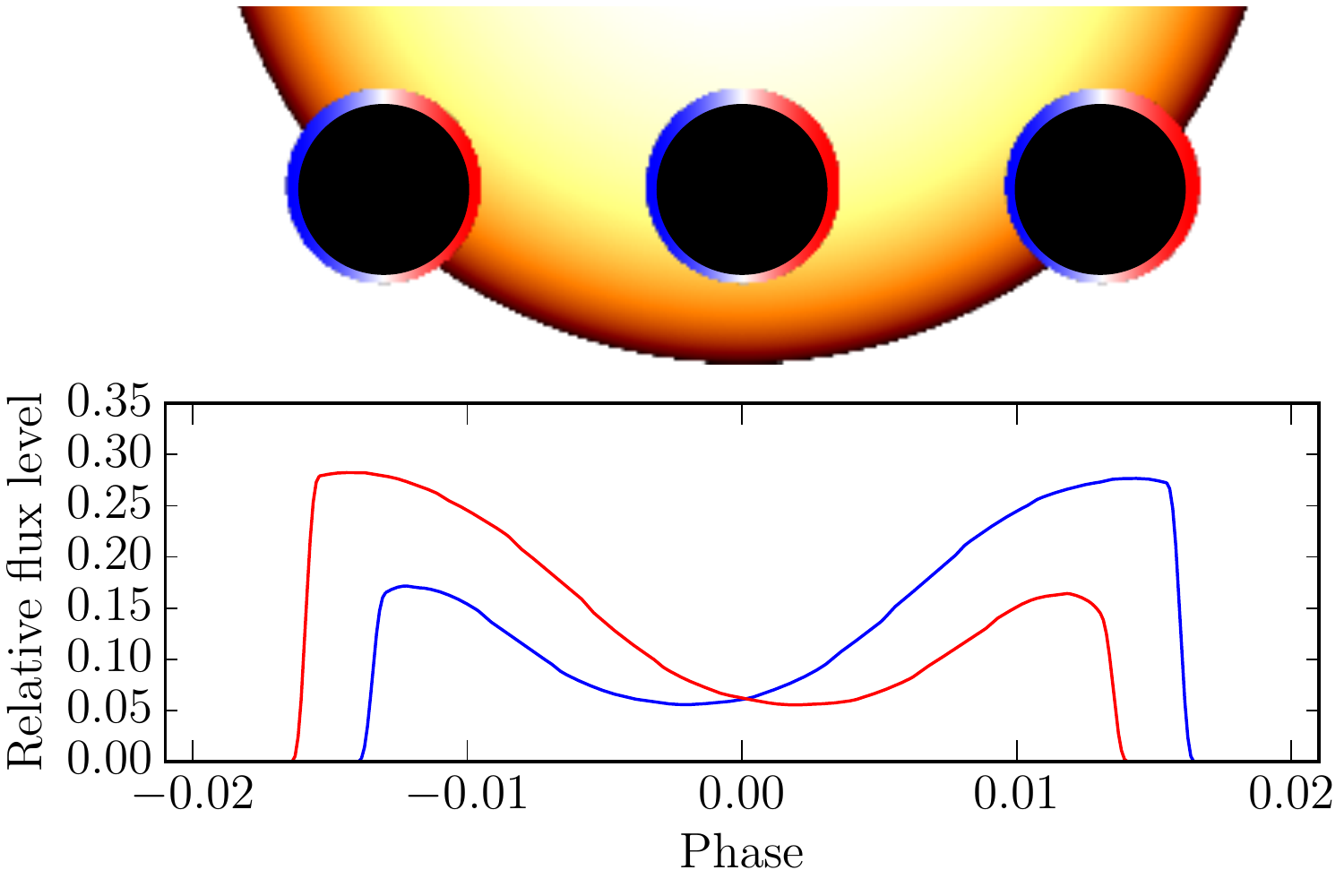}
\caption{\emph{Top Panel:} A scale diagram of the star-planet geometry, with the thickness of the planetary atmosphere doubled for clarity. The planet moves from left to right as it transits the star, with the leading limb coloured red and the trailing limb coloured blue. \emph{Bottom Panel:} Our model illustrates how the background illumination of the planetary limb changes during transit, allowing absorption by different parts of the planetary atmosphere to be separated. The red/blue lines represent the stellar flux available for sodium absorption at the planetary equator on the leading/trailing limbs (the model accounts for stellar rotation and limb darkening, and the planetary orbit). The asymmetry in illumination is strongest during ingress and egress, where absorption is also most heavily weighted (due to the relative Doppler shifts of the planetary and stellar sodium lines). }
\end{center}
\vspace{15pt}
\end{figure}

We search for differences in the Doppler shift of absorption from opposite sides of the planet by modeling the planetary absorption line profiles as a function of time through the transit. The power of our model to distinguish between atmospheric velocities at different points on the planetary limb is illustrated in Figure 1 where we compare the stellar flux available for absorption by atmospheric sodium on opposite sides of the planet. At each phase the stellar flux was sampled at the wavelengths corresponding to the sodium lines at the orbital velocity of the planet. During the transit, the orbital velocity of the planet causes the planetary lines to shift from blue to red by $\pm$ 16\,km\,s$^{-1}$, while the rotational velocity of the star as sampled by the planet varies from blue to red by $\pm$ 3\,km\,s$^{-1}$. The planetary sodium lines overtake the stellar lines at mid-transit, accounting for the relatively low weighting of this phase in Figure 1 (when the planetary and stellar velocity align there is little stellar flux for the planetary atmosphere to absorb). It is apparent from Figure 1 that with sufficiently high signal-to-noise data it is possible to model variations in the planetary absorption line profile to spatially resolve velocities in the planetary atmosphere.

The star-planet system was simulated on a 2D pixel grid, which implicitly accounts for both the atmospheric absorption and the Rossiter-McLaughlin effect. A quadratic limb darkening profile was used to give each pixel on the star a relative intensity \citep{Claret2011}. We adopted a stellar rotational velocity of $3.10 \pm 0.03$\,km\,s$^{-1}$, which was measured using Doppler tomography from the same set of HARPS observations \citep{Cameron2010}. Solid body rotation of the star was assumed in order to derive the rotation speed, but if this assumption is not correct the rotation velocity of the transit chord will not be affected. 

The absorption by the planet is modeled as an opaque occulting disk with the white light radius ratio of HD\,189733b. The atmospheric absorption is represented as an additional layer with absorption of the sodium doublet modeled by Voigt profiles. The atmospheric dynamics are parameterized by the equatorial velocities on the leading and trailing limbs of the planet together with the assumption of constant angular velocity (rigid body rotation). This allows continuous values of velocity to be defined at all grid points representing the planetary atmosphere with just two free parameters, while allowing higher velocities at the equator than the poles. This parameterization was selected in order to allow a reasonable approximation to the predictions of planetary rotation and an equatorial jet \citep{Showman2013, Miller2012}. The orbital velocity of the planet has been measured precisely from water \citep{Birkby2013} and carbon monoxide \citep{deKok2013, Rodler2013} absorption lines in the dayside spectrum of the planet, so the orbital velocity in the model is fixed at the measured value of 154 $\pm$ 4\,km\,s$^{-1}$. During transit the line-of-sight component of this orbital velocity is relatively small, varying from -16 to 16\,km\,s$^{-1}$, because the planet is moving almost perpendicular to our line of sight. 

The stellar line profile observed with HARPS is integrated over the entire disk of the star, and the natural line profile is broadened by macroturbulence and the rotational velocity of the star. The light blocked by the planet and its atmosphere as it transits the star will not be broadened by rotation, and this could potentially affect the transmission signal. To test for the importance of this effect, we attempted to recover the natural line profile of the sodium doublet through stellar atmosphere modeling. The Python package iSpec \citep{Blanco2011} was used to find a best fitting model for the stellar sodium absorption line. As the broadening parameters can be degenerate, we fixed the rotation velocity at the value of $3.10 \pm 0.03$\,km\,s$^{-1}$ \citep{Cameron2010}. The model was then fit for effective temperature, microturbulence, macroturbulence, surface gravity and limb darkening parameter. The best fitting model parameters were then used to synthesise a stellar profile that does not have the rotational broadening applied. In practice, we found that our final velocity measurements are not sensitive to the difference between this synthetic line profile and the integrated stellar line profile. 

\citet{Yan2015} find that the Fraunhofer lines of the Sun display a centre-to-limb variation which is not currently accounted for in our model, and \citet{Czesla2015} detect a similar variation in spectra of HD\,189733. Since the effect is primarily to alter the strength of the stellar lines, and the line variations will be symmetric on either side of the transit, we believe it is unlikely that this effect could lead to the asymmetric velocities we report in Section 4. However, further modelling of this effect will be the subject of future work.

Other effects that might distort the stellar line profile include convective blueshift, which has been noted to impact the Rossiter-McLaughlin effect \citep{Shporer2011}. However, the surface convective velocity for a K type star is only $\sim$200\,m\,s$^{-1}$, which causes an anomaly in the Rossiter-McLaughlin effect  at the $\sim$1\,m\,s$^{-1}$ level. Similarly, spots on the stellar surface are capable of mimicking spectral features in transmission spectra \citep{McCullough2014}, however, since the rotational velocity of the star is only $3.10$\,km\,s$^{-1}$, and a typical spot will not obscure more than a few percent of the surface, the effect on measured velocities should be no greater than a few 10s of m\,s$^{-1}$. The velocities for both convective blue shift and unocculted spots are substantially lower than the ones we report in Section 4.

We evaluate our model at the epoch of each HARPS spectrum using the position of the planet on the disk of the star calculated from the system parameters. The unbroadened stellar spectrum is Doppler shifted to the velocity of each of the grid points occulted by the planet. Planetary sodium absorption is calculated individually for each grid point after correcting for the orbital velocity and the modeled velocity of the planet atmosphere. The simulated absorption spectrum of the planet is normalised by a simulated out-of-transit spectrum to model the relative transmission. Our model accounts implicitly for the Rossiter-McLaughlin effect.

We attempted to spatially resolve the atmospheric velocities by applying our model simultaneously to all twenty in-transit spectra. We use a bootstrap analysis to assess the significance of the velocity measurements. By resampling with replacement from the set of twenty in-transit spectra many hundreds of times the analysis is robust to sources of systematic noise and individual outlier spectra.

\section{Results}\label{sec-disc}
Our average transmission spectrum and best fitting model is plotted in Figure 2, together with the stellar line profile and example planetary absorption line profiles at different transit phases showing the contributions of the opposite sides of the planet. The average transmission spectrum displays an asymmetric double peaked structure which is an artefact of the Rossiter-McLaughlin effect (accounted for in our model and discussed in Section 5).

\begin{figure}
\begin{center}
\includegraphics[width=1.0\linewidth]{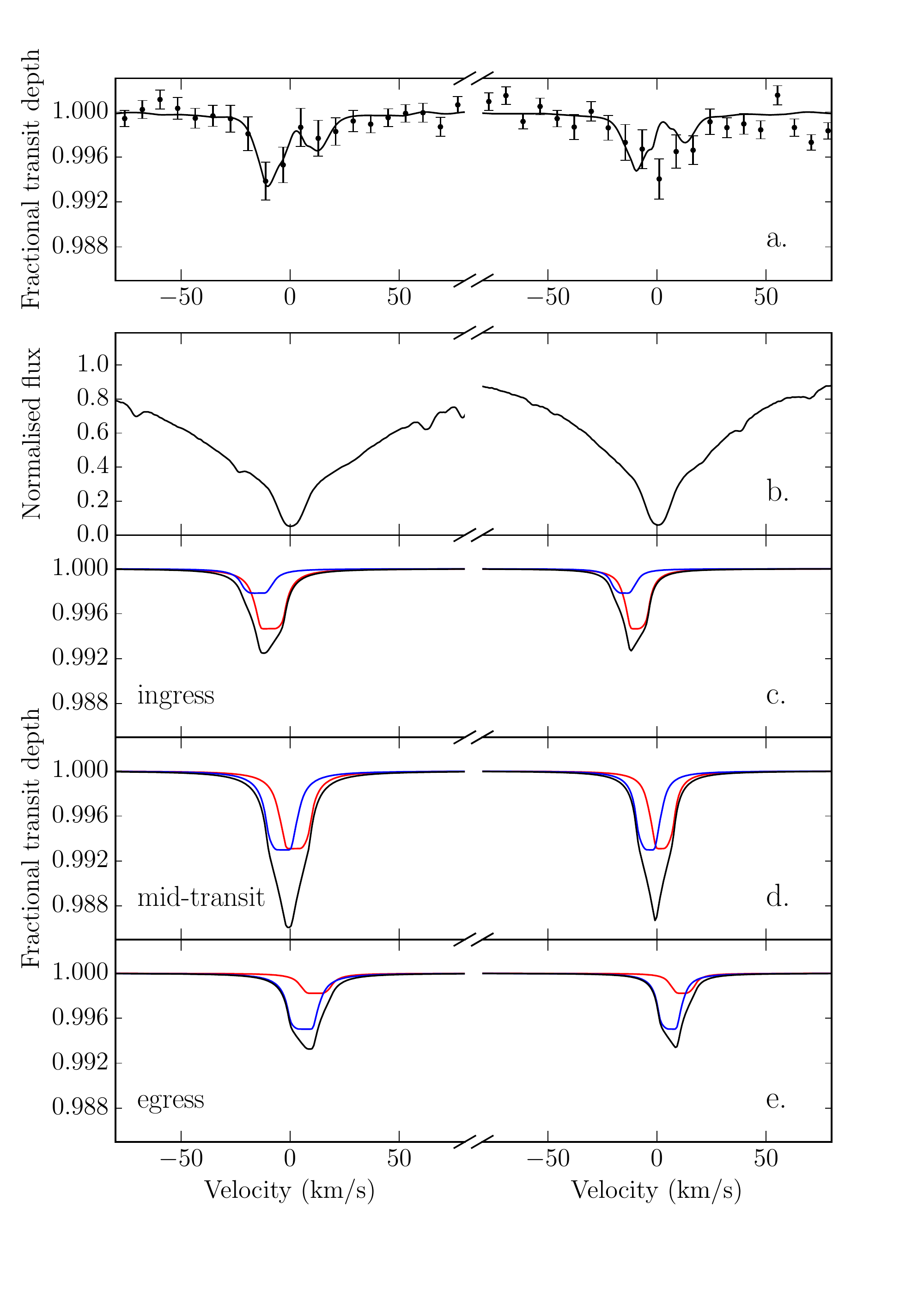}
\vspace{-30pt}
\caption{Planetary absorption line profiles. (a), the average sodium transmission spectrum of HD\,189733b calculated in the frame of the planet (binned by a factor of 10 for clarity) with the average of our time-dependent model overlaid. The asymmetric double-peaked line profile is caused by poor cancelation of the Rossiter-McLaughlin effect in the frame of the planet (accounted for implicitly in our model). (b), the stellar sodium line profile in the frame of the star. (c-e), our modeled planetary absorption line profiles in the stellar frame at ingress, mid-transit and egress respectively (with the Rossiter-McLaughlin effect removed for clarity). The contributions from the trailing (blue) and leading (red) hemispheres can be compared.}
\end{center}
\vspace{15pt}
\end{figure}

Our bootstrap distributions of atmospheric velocities are plotted in Figure 3, revealing a significant difference in the equatorial velocities on opposite sides of the planet. At the leading limb we measure a redshift of $2.3^{+1.3}_{-1.5}$\,km\,s$^{-1}$ and at the trailing limb we find a blue shift of $5.3^{+1.0}_{-1.4}$\,km\,s$^{-1}$. The two velocities differ by $7.6^{+2.0}_{-2.6}$\,km\,s$^{-1}$ and the bootstrap distributions differ with more than 99.7\% confidence. For the first time we have spatially resolved the atmospheric dynamics of an exoplanet. 

\begin{figure*}
\begin{center}
\includegraphics[width=1.0\linewidth]{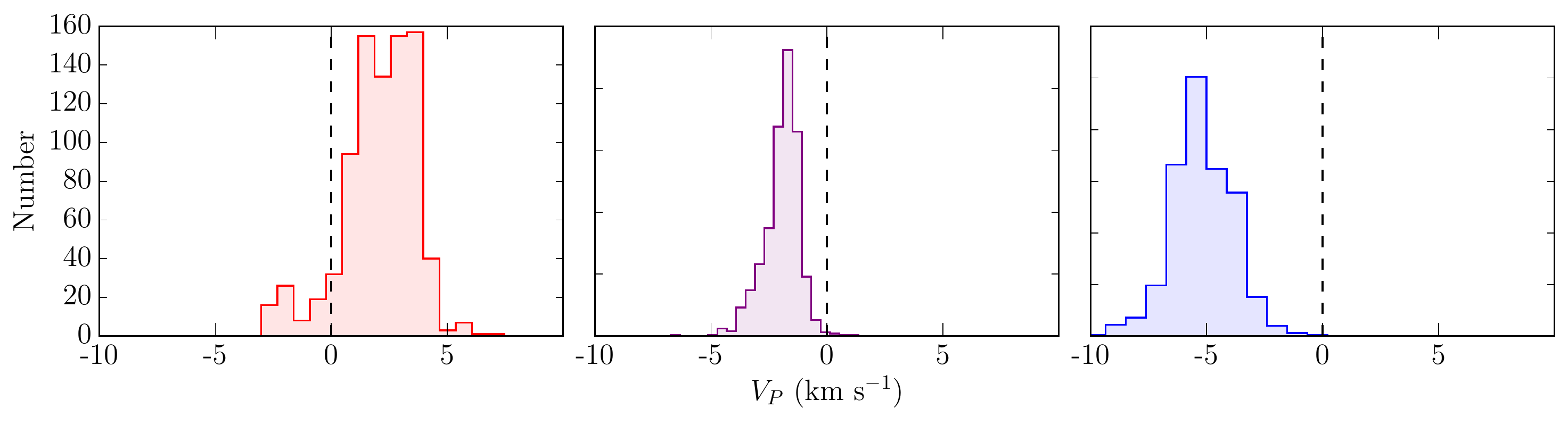}
\caption{The posterior distributions of atmospheric velocities from our bootstrap analysis. On the leading limb of the planet (left) a red shift of $2.3^{+1.3}_{-1.5}$\,km\,s$^{-1}$is found. The trailing limb is blue shifted by $5.3^{+1.0}_{-1.4}$\,km\,s$^{-1}$(right). The average velocity (middle) is found to be blue shifted with a velocity of $1.9^{+0.7}_{-0.6}$\,km\,s$^{-1}$. The strength and direction of the velocity offsets are consistent with a combination of tidally locked rotation and an eastward equatorial jet that is seen crossing from the dayside to the night side of the planet on the trailing limb.}
\end{center}
\vspace{15pt}
\end{figure*}

\begin{figure}
\begin{center}
\includegraphics[width=1.0\linewidth]{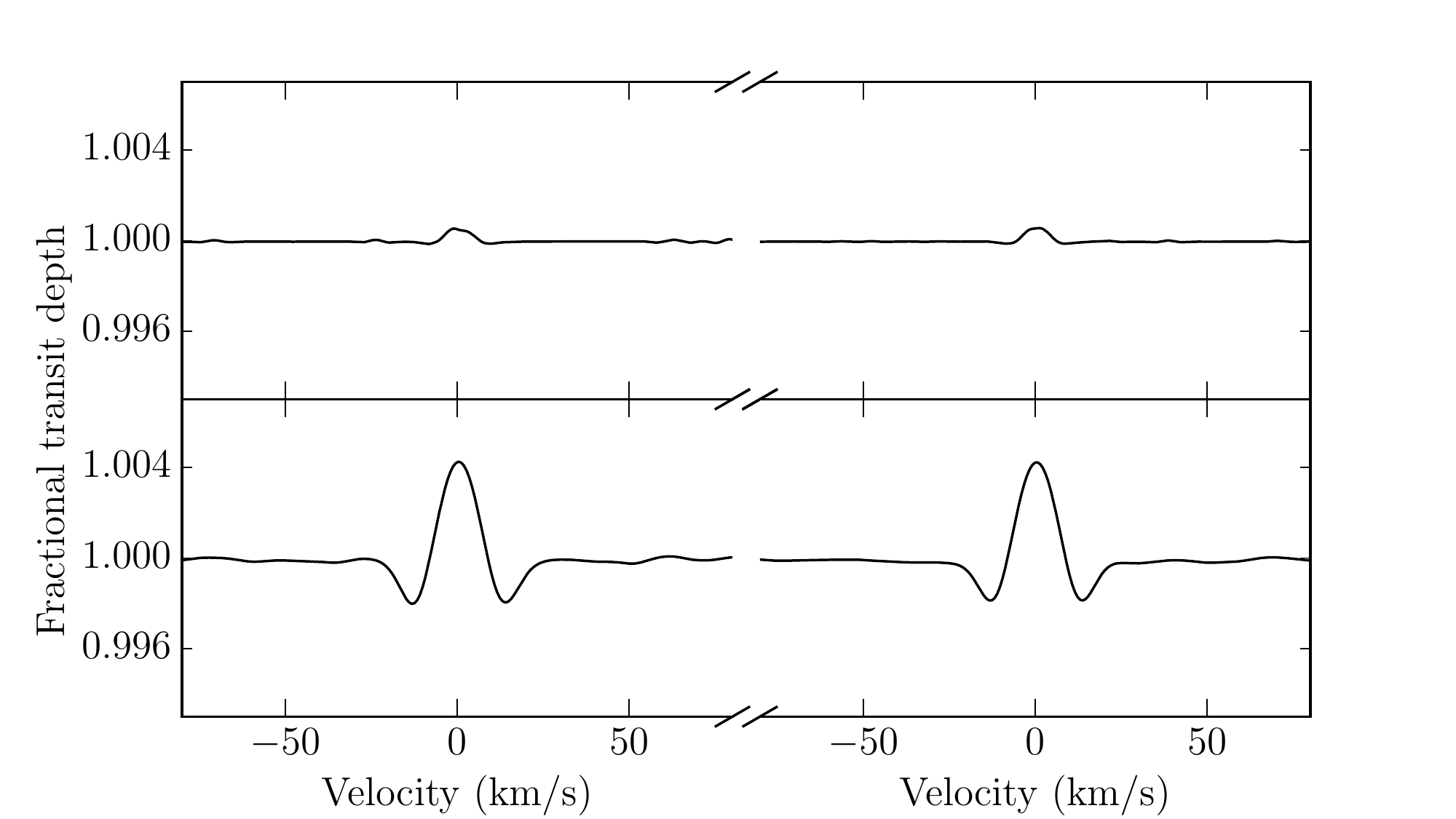}
\caption{Model calculations of transmission profile around the sodium doublet for a planet without an atmosphere to highlight the Rossiter-McLaughlin effect. \emph{Top Panel:} The transit spectra are aligned in the reference frame of the star. Here the Rossiter-McLaughlin effect is symmetric, and cancels except for a small residual caused by uneven phase coverage. \emph{Bottom Panel:} In this case the spectra are aligned in the planetary reference frame. This breaks the symmetry of the Rossiter-McLaughlin effect and leads to a spurious signal.}
\end{center}
\vspace{15pt}
\end{figure}

Assuming the rotation period of HD\,189733b is locked to its 2.22 day orbit we would expect symmetric red and blue shifts of 2.9\,km\,s$^{-1}$ on the opposite equators. Subtracting these expected rotational velocities from our measured equatorial velocities we find the velocity of the leading limb is consistent with tidally locked rotation (super-rotation of $-0.6^{+1.3}_{-1.5}$\,km\,s$^{-1}$), but that the trailing limb has an eastward excess velocity of $2.4^{+1.0}_{-1.4}$\,km\,s$^{-1}$ (68\% confidence errors). The bootstrap posterior distributions are not Gaussian, and we find that this super-rotation at the trailing limb is significant with 95\% confidence. 

\section{DISCUSSION and conclusions}\label{sec-disc}

Our results show that the atmosphere of HD\,189733b has a strong eastward motion that is the same direction and has a similar amplitude to that expected from tidally locked planetary rotation. For a close-in planet such as HD\,189733b it is almost inevitable that tidal forces will synchronise the rotation to the orbital period in just a few million years \citep{Guillot1996}.

The super-rotating material we see absorbing at the trailing limb of HD\,189733b is separated by a longitude of only 60\degree from the atmospheric hotspot observed to be offset eastward from the substellar point with Spitzer \citep{Knutson2007, Knutson2008}. It is likely, therefore, that our measured velocity excess is a direct detection of the eastward equatorial jet that is predicted in atmospheric circulation models \citep{Showman2002} and has been invoked to explain the offset of the atmospheric hotspot in HD\,189733b. The observed jet velocity and the lower velocity on the leading limb of the planet are remarkably consistent with predictions from state-of-the-art circulation models for HD\,189733b \citep{Showman2013}.

Our detection of this excess velocity is in line with the previous detections of net blue shift in the average transmission spectra of HD\,209458b \citep{Snellen2010} and HD\,189733b \citep{Wyttenbach2015}. However, by spatially resolving the atmospheric dynamics we have been able to locate the excess blue shift to the trailing limb of the planet. For comparison with the unresolved studies, our limb-averaged velocity offset is $1.9^{+0.7}_{-0.6}$\,km\,s$^{-1}$ (Figure 3), which is consistent with the average velocity offset of carbon monoxide absorption measured for HD\,209458b \citep{Snellen2010}, but significantly smaller than the value of 8 $\pm$ 2\,km\,s$^{-1}$ found for sodium in HD\,189733b \citep{Wyttenbach2015}. We believe this discrepancy is due to the Rossiter-McLaughlin effect, which is accounted for implicitly in our time-resolved model.

The Rossiter-McLaughlin effect is usually assumed to be symmetric and to cancel out in average transmission spectra. This is true when transmission spectra are constructed in the frame of the star, however, in the frame of the planet the symmetry is broken and the Rossiter-McLaughlin effect causes a spurious signal. This is illustrated in Figure 4, where the effect on transmission spectra when analysed in both the reference frame of the star and of the planet is shown. In the stellar frame the Rossiter-McLaughlin effect cancels except for residuals caused by slightly uneven phase coverage in our data. In the planetary reference frame the effect no longer cancels and there is a strong artefact, even with ideal phase coverage. This is accounted for implicitly in our model and is apparent in the asymmetric double-peaked line profile of our best-fitting model in Figure 2. Fitting this asymmetric line profile with a symmetric model recovers the larger (spurious) velocity offset found by \citet{Wyttenbach2015}, who analyse their spectra in the reference frame of the planet. We note that even in the stellar frame the Rossiter-McLaughlin effect only cancels if the phase coverage is symmetric, otherwise leading to large systematics that likely account for the very high blue shifted velocity of $37$\,km\,s$^{-1}$ reported by \citet{Redfield2008} and \citet{Jensen2011} for sodium absorption in HD\,189733b. Since our model is evaluated at the specific epochs of the individual HARPS spectra, the phase coverage is accounted for implicitly and does not affect our fitted velocities.

We note further that \citet{Brogi2015} have analysed high resolution infra-red spectroscopy of CO and H$_{2}$O absorption in HD\,189733b and find a global blue shift of $-1.7^{+1.1}_{-1.2}$\,km\,s$^{-1}$ which is consistent with our disc averaged measurement.

Our spatially resolved velocities show that absorption by the planetary atmosphere is intrinsically velocity broadened, suggesting that previous attempts to model the line profiles in terms of atmospheric temperature and pressure are likely to have over-estimated the role of thermal Doppler broadening \citep{Huitson2012, Wyttenbach2015, Heng2015}. Similarly, the extent of the planetary atmosphere inferred from the depth of the average line will tend to have been under-estimated \citep{Redfield2008, Jensen2011, Huitson2012, Wyttenbach2015}.

As can be seen in Figure 1, the asymmetry in background illumination extends across the planetary limb. With sufficiently high signal-to-noise observations it should be possible to extend our model to recover wind maps at higher spatial resolution.

\acknowledgements{

T.L. is supported by a STFC studentship. P.W. is supported by a STFC consolidated grant (ST/L000733/1). Based on data products from observations made with ESO Telescopes at the La Silla Paranal Observatory under programme ID 079.C-0127(A).}

\end{document}